\documentclass[fleqn,twoside]{article}
\usepackage{epsfig,espcrc2}

\readRCS
$Id: espcrc2.tex,v 1.2 2004/02/24 11:22:11 spepping Exp $
\usepackage{graphicx}
\usepackage[figuresright]{rotating}
\input defs.tex

\newcommand{\AmS}{{\protect\the\textfont2
  A\kern-.1667em\lower.5ex\hbox{M}\kern-.125emS}}
\title{\center{Determining the unitarity triangle angle $\gamma$ with a four-body amplitude analysis of \BbtoD2k2pK decays}}

\author{J. Rademacker$^1$, G. Wilkinson \\ \vspace{0.2cm}
University of Oxford, Denys 
Wilkinson Building, Keble Road, Oxford, OX1 3RH, United Kingdom. \\ 
\vspace*{0.1cm}
$^1$ Now at: University of Bristol, H.H. Wills 
Physics Laboratory, Tyndall Avenue, Bristol, BSS 1TL, United Kingdom.}

\begin{document}

\begin{abstract}
We explain how a four-body amplitude analysis of the $\rm D$ decay
products in the mode \BbtoD2k2pK is sensitive to the unitarity triangle
angle $\gamma$.  We present results from simulation studies which show
that a precision on $\gamma$ of $15^\circ$ is achievable with 1000
events and assuming a value of $0.10$ for the parameter $r_B$. 
\end{abstract}

\maketitle

\section{Introduction}

A precise measurement of the unitarity triangle angle $\gamma$ is one
of the most important goals of CP violation experiments.
$\gamma$ is defined 
as arg$(-V^\ast_{ub}V_{ud}/V_{cb}^\ast V_{cd})$, where $V_{ij}$ are the 
elements of the Cabibbo-Kobayashi-Maskawa (CKM) mixing matrix. In the
Wolfenstein convention~\cite{WOLF} $\gamma =$~arg$(V_{ub}^\ast)$.

\begin{figure}[htb]
\begin{center}
\epsfig{file=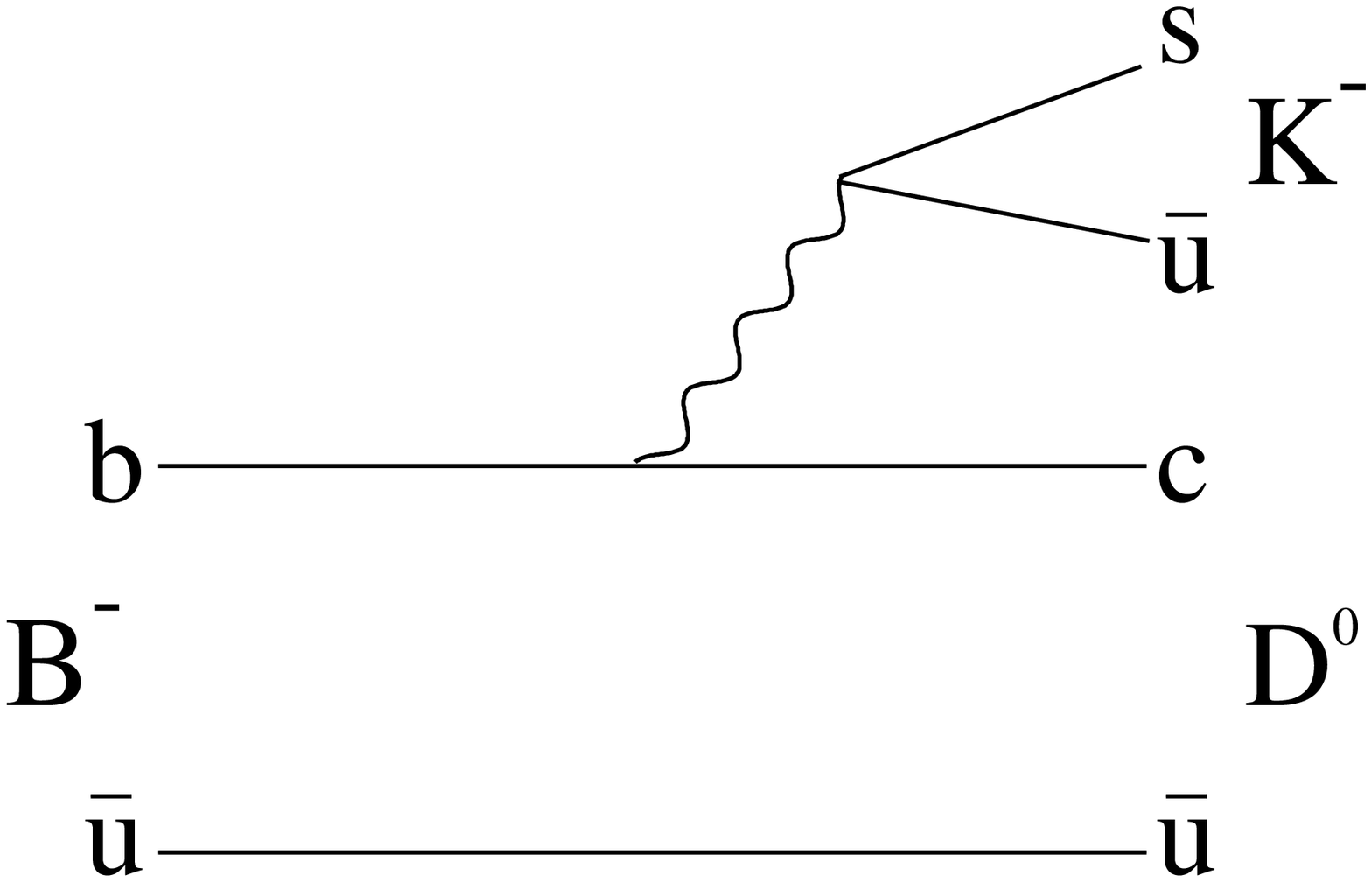,width=0.43\textwidth}
\epsfig{file=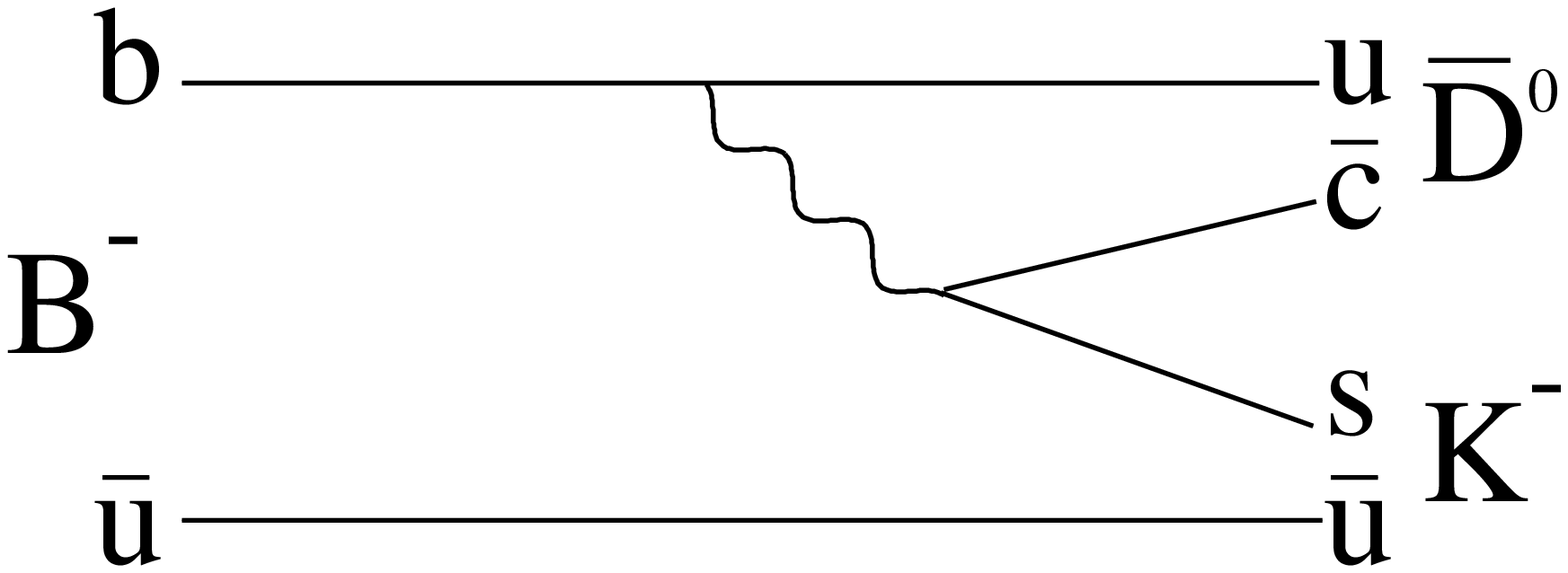,width=0.43\textwidth}
\end{center}
\vspace*{-1.0cm}
\caption[]{The diagrams for $\rm B^- \to D^0 K^-$ and $\rm B^- \to 
\bar{D}^0 K^-$.   There is a relative phase of $\delta_B \,-\,\gamma$
between the two amplitudes, and a relative magnitude of $r_B$.}
\label{fig:diagrams}
\end{figure}

A class of promising methods to measure $\gamma$ exists which exploits
the interference between the amplitudes leading to the decays 
\BmtoDK and \BmtoDbK (Figure~\ref{fig:diagrams}), where the  $\rm D^0$ and  
$\rm \bar{D}^0$ are reconstructed in a common final state.
This final state may be, for example, a CP eigenstate such
as $\rm K^+K^-$ (`GLW method')~\cite{GLW}, or a non-CP eigenstate such as $\rm K^+\pi^-$, 
which can be reached both through a doubly Cabibbo-suppressed $\rm D^0$ decay 
and a Cabibbo-favoured  $\rm \bar{D}^0$ decay (`ADS method')~\cite{ADS}.
Recent attention has focused on  self-conjugate three-body final
states, in particular  $\rm D \to K_S\pi^+\pi^-$ \footnote{Here and subsequently $\rm D$ 
signifies either a $\rm D^0$ or a $\rm \bar{D}^0$.}.
Here a Dalitz analysis of the resonant substructure in the $\rm K_S \pi^+\pi^-$
system allows $\gamma$ to be extracted~\cite{DALITZTH2}.
The B-factory experiments have 
used this method
to obtain the first interesting direct constraints 
on $\gamma$~\cite{DALITZBABAR,DALITZBELLE}.

Here we explore the potential of determining $\gamma$ through
a four-body amplitude analysis of the $\rm D$ decay products 
in the mode \BbtoD2k2pK .  CP studies involving
amplitude analyses of four-body systems have been proposed 
elsewhere~\cite{DALITZTH2}, and strategies already exist for \BgtoDK 
approaches exploiting singly-Cabibbo suppressed decays~\cite{SCS}.
Our method benefits from a final state that involves only charged
particles, which makes it particularly suitable for experiments 
at hadron colliders, most notably LHCb.

This paper is organised as follows.  In Section~\ref{sec:basics} we summarise
the essential features of \BgtoDK decays, and state the present knowledge
of the parameters involved, and of the decay \Dgto2k2p .
In Section~\ref{sec:meat} 
a full model of \BbtoD2k2pK decays is developed, which is then used within
a simulation study to estimate the precision on $\gamma$ 
which may be obtained through a four-body amplitude analysis. We conclude in 
Section~\ref{sec:conc}.

\section{\BgtoDK decays and \Dgto2k2p }
\label{sec:basics}


Let us define the amplitudes of the two diagrams illustrated in 
Figure~\ref{fig:diagrams} as follows
\begin{eqnarray}
A(\BmtoDK) & \equiv & A_B , \\
A(\BmtoDbK) & \equiv & A_B r_B e^{i(\delta_B - \gamma)}.
\end{eqnarray}
Here the strong phase of $A_B$ is set to zero by convention, and $\delta_B$ is
the difference of strong phases between the two amplitudes. $\gamma$ represents
the weak phase difference between the amplitudes, where contributions to the 
CKM elements of order $\lambda^4$ and higher 
(with $\lambda$ being the sine of the 
Cabibbo angle) have been neglected. In the CP-conjugate 
transitions $\gamma \to -\gamma$, whereas $\delta_B$ remains unchanged.
$r_B$ is the relative magnitude of
the colour-suppressed \BmtoDbK process to the colour-favoured \BmtoDK
transition.
Preliminary indications as to the values of $\gamma$, $\delta_B$ and $r_B$
come from the   $\rm B^\pm \to D K^\pm, D \to K_S \pi^+\pi^-$ analyses
performed at the B-factories~\cite{DALITZBABAR,DALITZBELLE}. 
Fits to the ensemble of hadronic flavour data also provide indirect 
constraints on the value of $\gamma$~\cite{CKMFIT,UTFIT}.
These
results lead us to assume values of $\gamma=60^\circ$ and
$\delta_B=130^\circ$ for the illustrative sensitivity studies
presented in Section~\ref{sec:meat}.  We set $r_B$  to $0.10$,
which is the approximate average of the Dalitz results  
and the lower values favoured by the ADS and 
GLW analyses~\cite{ADSGLWBABAR,ADSGLWBELLE}.

Results have recently been reported from an amplitude analysis of the 
decay \Dgto2k2p \cite{FOCUS}, which shows that the dominant contributions
come from $\rm D \to AP$ and $\rm D \to VV$ modes. 
Earlier measurements of \Dgto2k2p were published in~\cite{E791}.
Our sensitivity studies for the $\gamma$ extraction, 
presented in Section~\ref{sec:meat}, 
are based on the results found in~\cite{FOCUS}.

The branching ratio of the mode \BbtoD2k2pK can be estimated
as the product of the two meson decays,
and found to be $9.2 \times 10^{-7}$~\cite{PDG}.  
This channel is particularly well 
matched  to the  LHCb experiment, on account of 
the kaon-pion  discrimination provided by the RICH system, and the 
absence of any neutrals in the final state, which allows for good
reconstruction efficiency and powerful vertex constraints.
Consideration of the trigger and reconstruction efficiencies of similar
topology decays reported in~\cite{LHCBLITE} leads to the expectation of 
sample sizes of more than 1000 events per year of operation.

\section{Estimating the $\gamma$ sensitivity in \\ \BbtoD2k2pK decays}
\label{sec:meat}

In this section we formulate a model to describe \BbtoD2k2pK decays.
This model neglects $\rm D^0 - \bar{D}^0$ oscillations and CP violation
in the $\rm D$ system, which is a good approximation in the Standard Model.
The model is then used in a simulation study to estimate the sensitivity
with which $\gamma$ can be determined from an analysis of
\BbtoD2k2pK events.  

\subsection{Decay Model}
\label{sec:model}

 In the same way as the kinematics of a three-body decay can be fully
 described by two variables (Dalitz Plot), typically \( s_{12} = (p_1
 + p_2)^2 \), \( s_{23} = (p_2 + p_3)^2 \), where $p_1, p_2, p_3$ are
 the 4-momenta of the final state particles, so can a four-body decay be
 described by five variables. In this paper we use the following
 convention for labelling the particles involved in the D decay
and their 4-momenta:
\begin{flushleft}
\begin{tabular}{lcccccc}
 Decay: &
\prt{D} & $\to$ & \prt{K^+} & \prt{K^-} & \prt{\pi^+} & \prt{\pi^-}, \\
Label: &
 0 & & 1& 2& 3& 4 ,
\\
4-mom. : & $p_0$& & $p_1$& $p_2$& $p_3$& $p_4$ .
\end{tabular}
\end{flushleft}
\noindent We also define:
\begin{eqnarray}
 s_{ij} & \equiv & (p_i + p_j)^2 \, , \nonumber \\
 s_{ijk} & \equiv & (p_i + p_j + p_k)^2 \, , \nonumber \\  
 t_{ij} & \equiv & (p_i - p_j)^2 \, .
\end{eqnarray}
\noindent We then choose a set of five variables to describe the decay kinematics:
$t_{01}=s_{234}$, $s_{12}$, $s_{23}$, $s_{34}$ and $t_{40}=s_{123}$.
From these variables all other invariant masses $s_{ij}, s_{ijk}$, and, for a
given frame of reference, all momenta $p_i$ can be calculated.

In contrast to the phase space density for three-body decays, which is
uniform in terms of the usual parameters $s_{12}, s_{23}$, four-body
phase space density, $d\phi/d t_{01} d s_{12} d s_{23} d s_{34} d t_{40}$,
is not flat in 5 dimensions, but proportional to
the square-root of the inverse of the 4-dimensional Grahm determinant
\cite{BYCKLING_KAJANTIE}:
\begin{eqnarray}
\lefteqn{
\frac{ d \phi
     }{
       d t_{01} d s_{12} d s_{23} d s_{34} d t_{40}
     }
=}&&
\nonumber\\
&&
\frac{\pi^2}{32 m_0^2}
 \left( -
\left|
   \begin{array}{cccc}
    x_{11} &  x_{12} &  x_{13} & x_{14} \\
    x_{21} &  x_{22} &  x_{23} & x_{24} \\
    x_{31} &  x_{32} &  x_{33} & x_{34} \\
    x_{41} &  x_{42} &  x_{43} & x_{44} \\
\end{array}
\right|
 \right)^{-\frac{1}{2}},
\end{eqnarray}
where
$x_{ij} \equiv p_i \cdot p_j = \frac{1}{2}\left(s_{ij} - m_i^2 - m_j^2 
\right)$.

The total decay amplitude for the \prt{D^0} decay to the
$\rm K^+K^-\pi^+\pi^-$ final state
is the sum over all individual amplitudes $A_k$ to each
set of intermediate states $k$, weighted by a complex factor 
$|c_k| e^{i\phi_k}$
\begin{equation}
A_{D^0} = \sum_k |c_k| e^{i\phi_k} A_k \, .
\end{equation}
An analysis of the $\rm D^0 \to K^+K^-\pi^+\pi^-$ decay 
amplitude is reported in~\cite{FOCUS}, which
fits 10 separate contributions.
In this analysis, however, no distinction is made
between the modes $\rm D^0 \to K_1(1270)^+K^-$,
$\rm K_1(1400)^-K^+$ and $\rm K^*(892)^0 K^-\pi^+$,
and those decays to the CP-conjugate final states.
In our study we base $|c_k|$ and $\phi_k$ on the values 
found in~\cite{FOCUS}, but consider different scenarios
for the relative contributions of the above modes.  
In order to label these scenarios we make the definitions
\begin{eqnarray}
R_{K_1(1270)K} &\equiv&  \frac{|c_{K_1(1270)^+K^-}|^2}{|c_{K_1(1270)^-K^+}|^2}
\label{eq:Rdef}
\end{eqnarray}
\noindent and
\begin{eqnarray}
\Delta \phi_{K_1(1270)K} &\equiv& \phi_{K_1(1270)^+K^-}  \nonumber \\
                         &   -  & \phi_{K_1(1270)^-K^+}.
\label{eq:Dpdef}
\end{eqnarray}
\noindent We define similar variables for the $\rm D^0 \to K_1(1400)^\pm K^\mp$
and $\rm D^0 \to K^*(892)^0 K^\pm \pi^\mp$ decays.  Our default scenario
assumes the arbitrary values 
$R_{K_1(1270)K}=R_{K_1(1400)K}=R_{ K^*(892)^0 K \pi }=1$,
$\Delta \phi_{K_1(1270)K}=39^\circ $, $\Delta \phi_{K_1(1400)K}=211^\circ $ and
$\Delta \phi_{ K^*(892)^0 K \pi }=115^\circ$.


The amplitudes $A_k$ are constructed as a product of form
factors ($F_l$), relativistic Breit-Wigner functions ($BW$), 
and spin amplitudes ($s_l$) which account for angular 
momentum conservation, where $l$ is the angular 
momentum of the decay vertex.   Therefore  the decay amplitude
with a single resonance is given by
\begin{equation}
A = F_{l} \cdot s_l \cdot BW \, ,
\end{equation}
(where the subscript $k$ has now been omitted), and
a decay amplitude with two resonances $\alpha$ and $\beta$ is written
\begin{equation}
A = s_l \cdot F_{l\;\alpha} \cdot BW_{\alpha}
        \cdot F_{l\;\beta}  \cdot BW_{\beta}.
\end{equation}
For $F_l$ we use Blatt-Weisskopf damping factors~\cite{BLATTWEISSKOPF}
and for $s_l$ we use the Lorentz invariant amplitudes~\cite{MARKIII}, 
which depend both on the spin of the resonance(s) and the 
orbital angular momentum. 

With these definitions, the total decay amplitude for
\prt{B^{-} \to D K^-, D 
 \to K^+K^-\pi^+\pi^-} is given by
\begin{eqnarray}
 A^- &=&  A\left(
 \prt{B^{-} \to \left(K^+K^-\pi^+\pi^-\right)_{D} K^-}
 \right)
\nonumber\\
 &=& A_B \left( A_{D^0} + 
   r_B e^{i\left(\delta_B - \gamma\right)}
   \overline{A_{D^0}} \right)
\nonumber\\
 &=& A_B \Big(A_{D^0}\left( t_{01}, s_{12}, s_{23}, s_{34}, t_{40} \right)
\nonumber\\
 & &  
   + \, r_B e^{i\left(\delta_B - \gamma\right)} \cdot \nonumber\\
 & &  A_{D^0}\left( t_{02}, s_{12}, s_{14}, s_{34}, t_{30} \right) \Big).
\end{eqnarray}
 The corresponding expression for the CP conjugate decay is
\begin{eqnarray}
A^+ &=& A\left(
 \prt{B^{+} \to \left(K^+K^-\pi^+\pi^-\right)_{D} K^+}
 \right)
\nonumber\\
 &=& A_B \left( \overline{A_{D^0}} + 
   r_B e^{i\left(\delta_B + \gamma\right)}
   A_{D^0} \right)
\nonumber\\
 &=& A_B \Big(A_{D^0}\left( t_{02}, s_{12}, s_{14}, s_{34}, t_{30} \right)
\nonumber\\
 & & + \,
   r_B e^{i\left(\delta_B + \gamma\right)} \cdot \nonumber \\
& &   A_{D^0}\left( t_{01}, s_{12}, s_{23}, s_{34}, t_{40} \right) \Big).
\end{eqnarray}

The total probability density function for a \prt{B^{-} \to
\left(K^+K^-\pi^+\pi^-\right)_{D} K^-} event is then given by

\begin{equation}
 P^- = N |A^-|^2
\frac{ d \phi
     }{
       d t_{01} d s_{12} d s_{23} d s_{34} d t_{40}
     },
\label{eq:pdf}
\end{equation}
(with an equivalent expression for \prt{B^+} decays) 
 where $N$ is an appropriate normalisation factor which
may be obtained through numerical integration.

\subsection{Simulation Study}

\begin{figure}
\begin{center}
\begin{tabular}{ll}
\epsfig{file=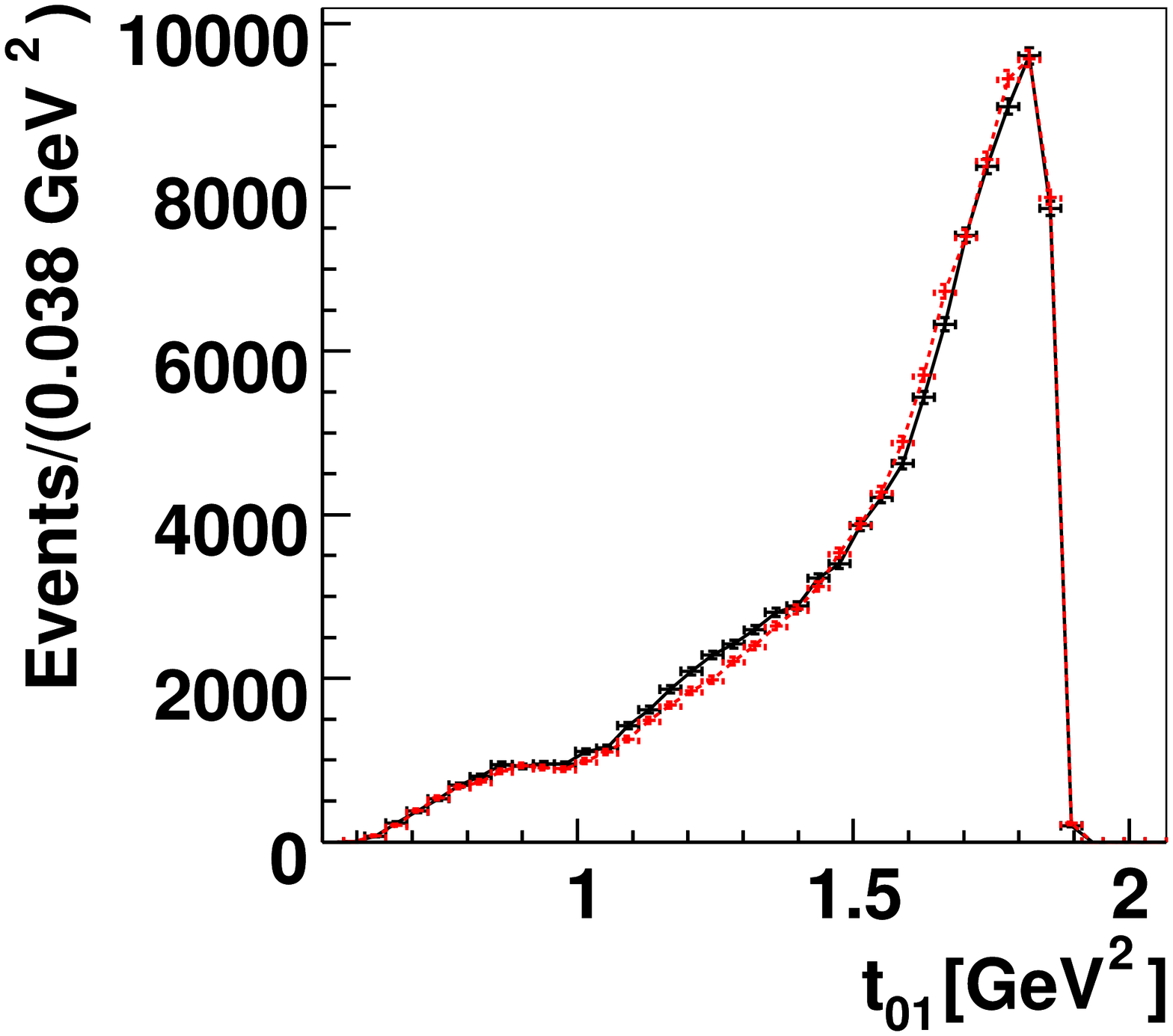,width=0.23\textwidth} & \epsfig{file=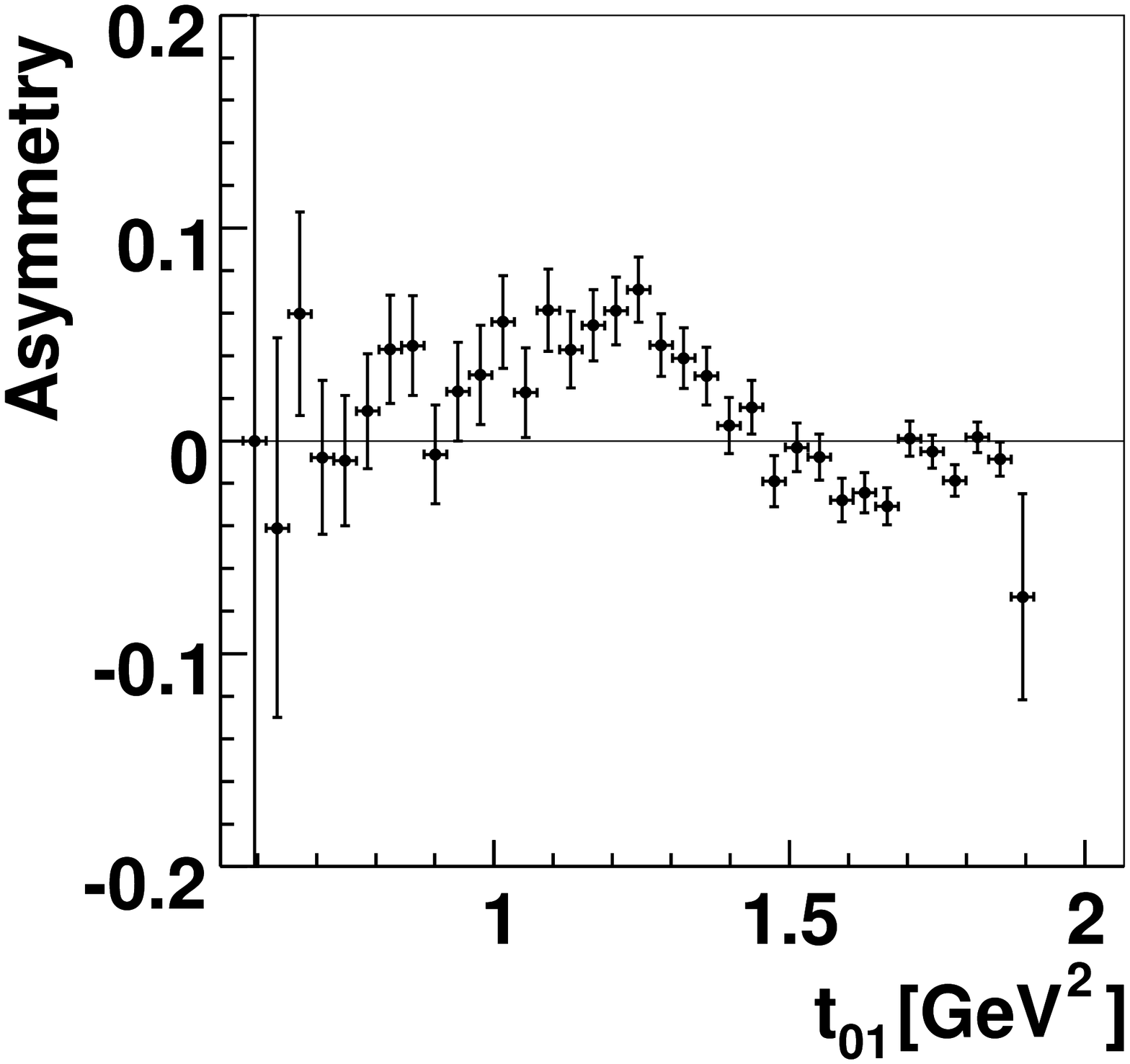,width=0.215\textwidth} \\
\epsfig{file=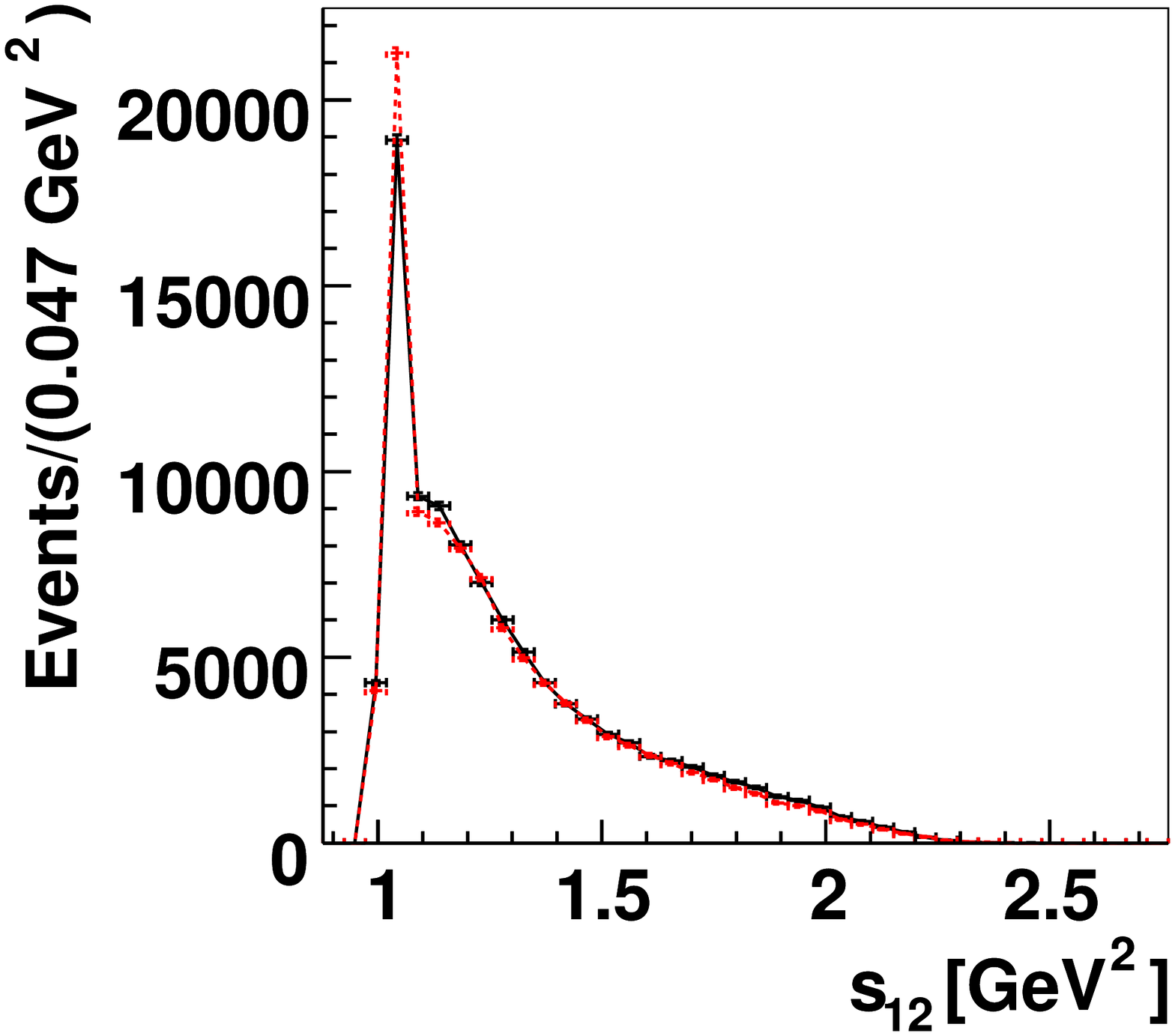,width=0.23\textwidth} & \epsfig{file=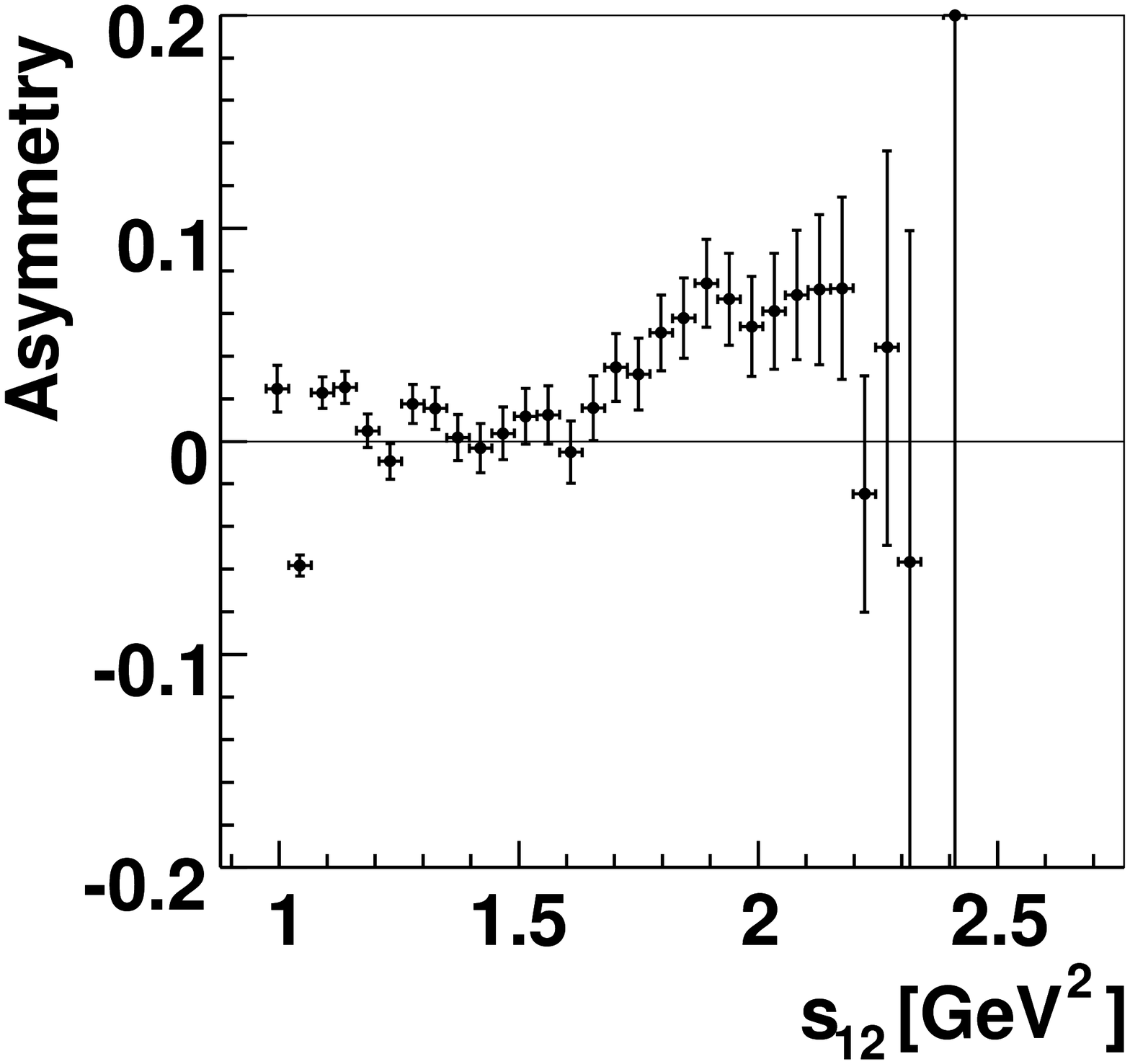,width=0.215\textwidth} \\
\epsfig{file=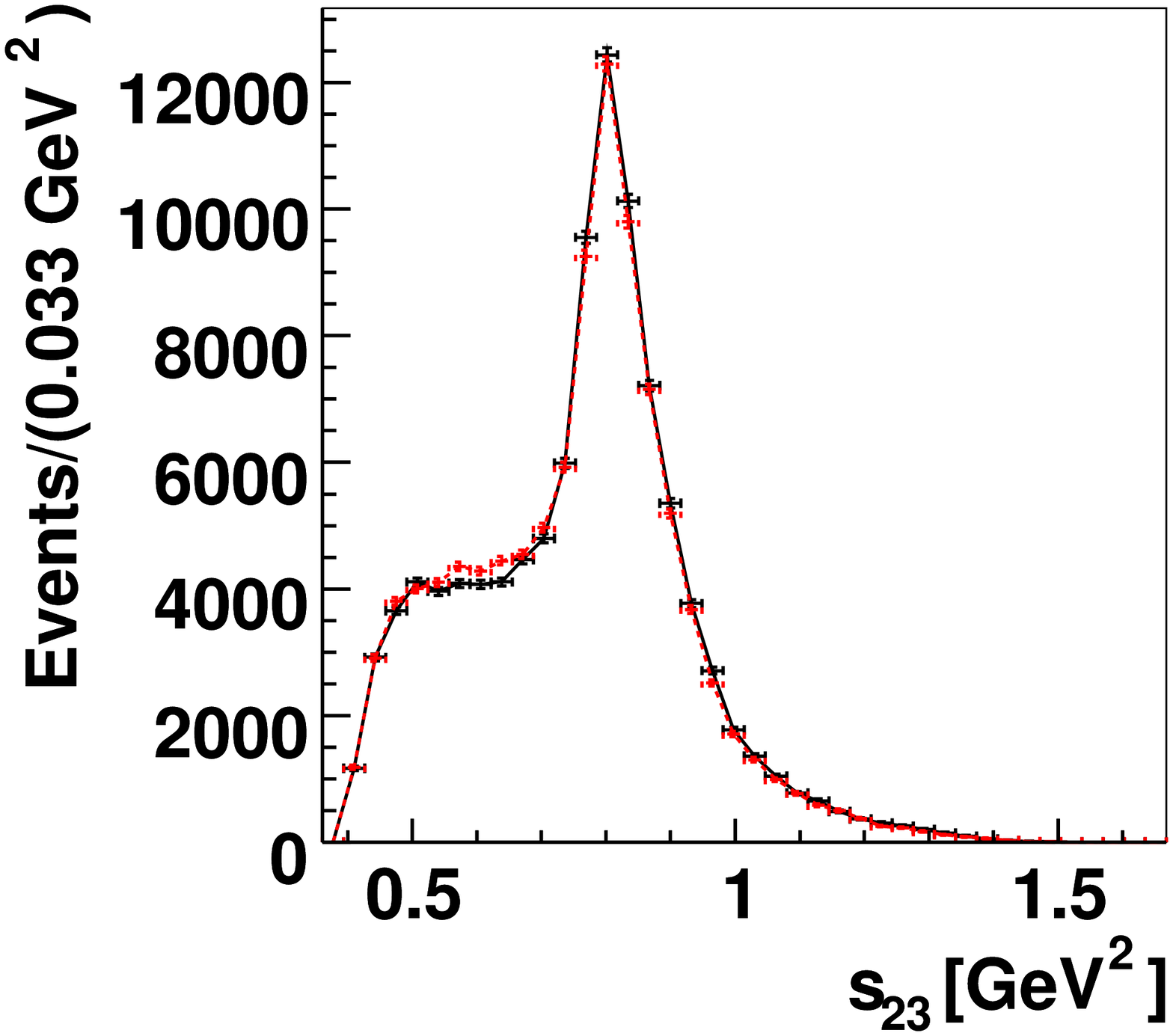,width=0.23\textwidth} & \epsfig{file=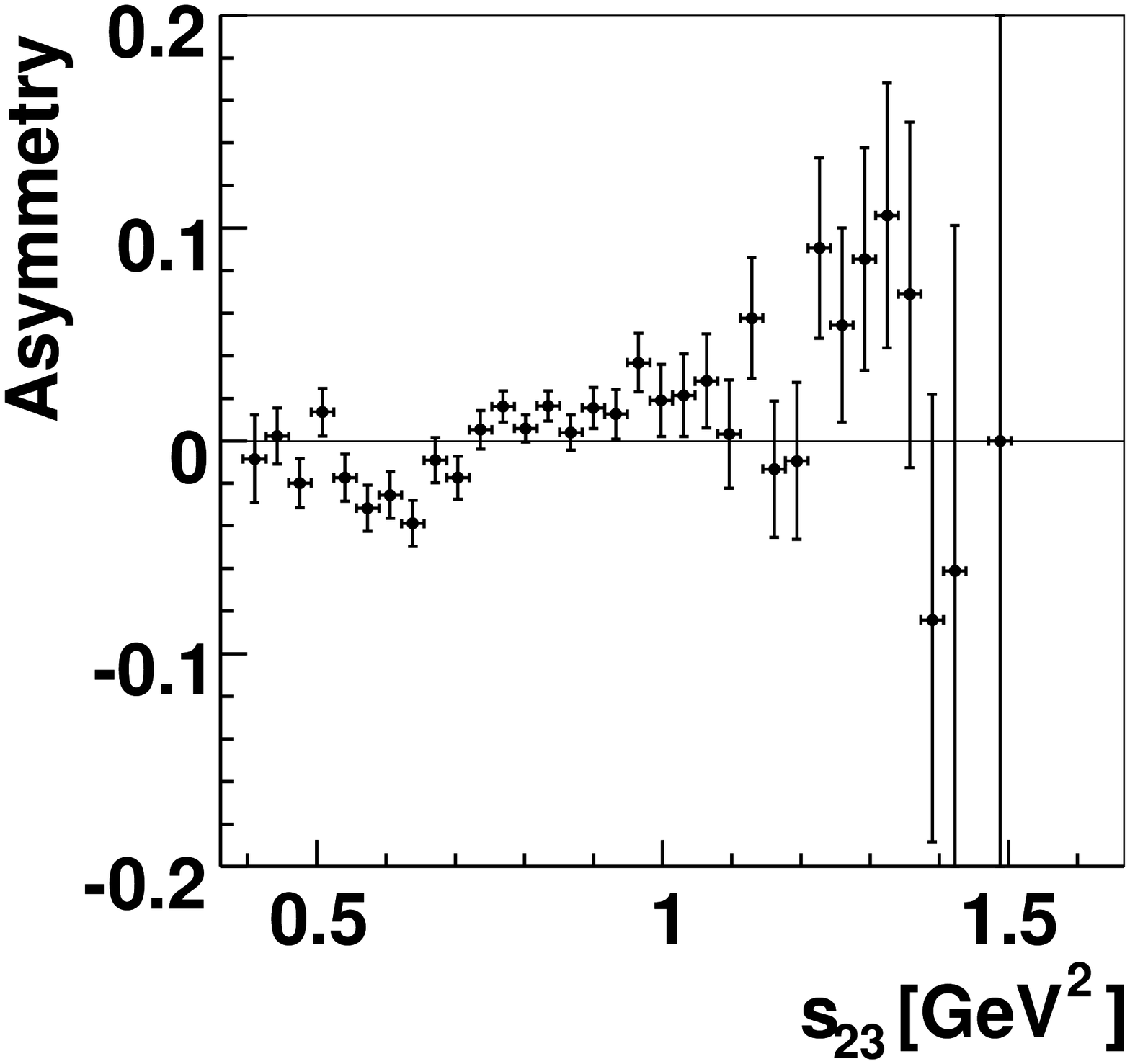,width=0.215\textwidth} \\
\epsfig{file=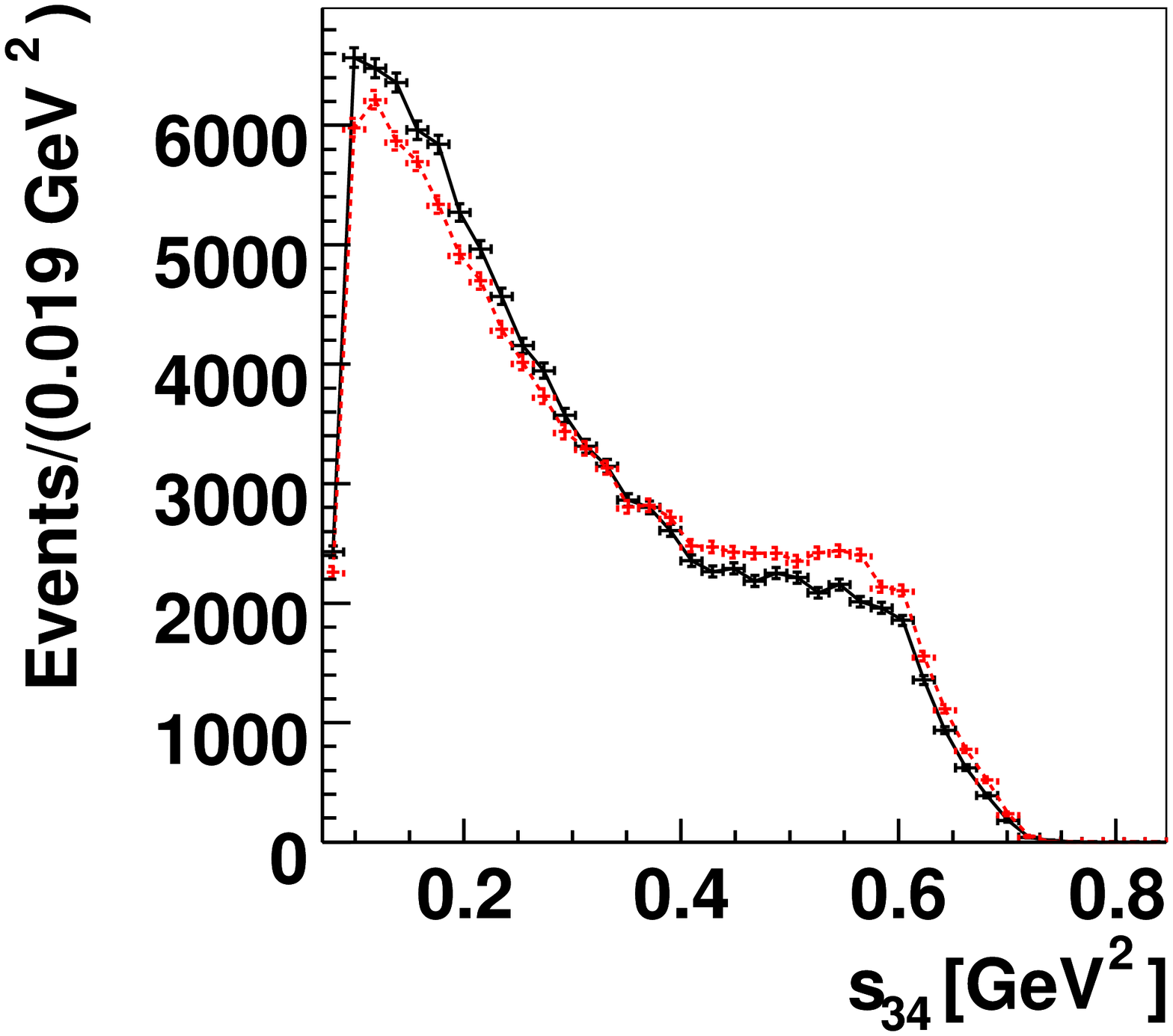,width=0.23\textwidth} & \epsfig{file=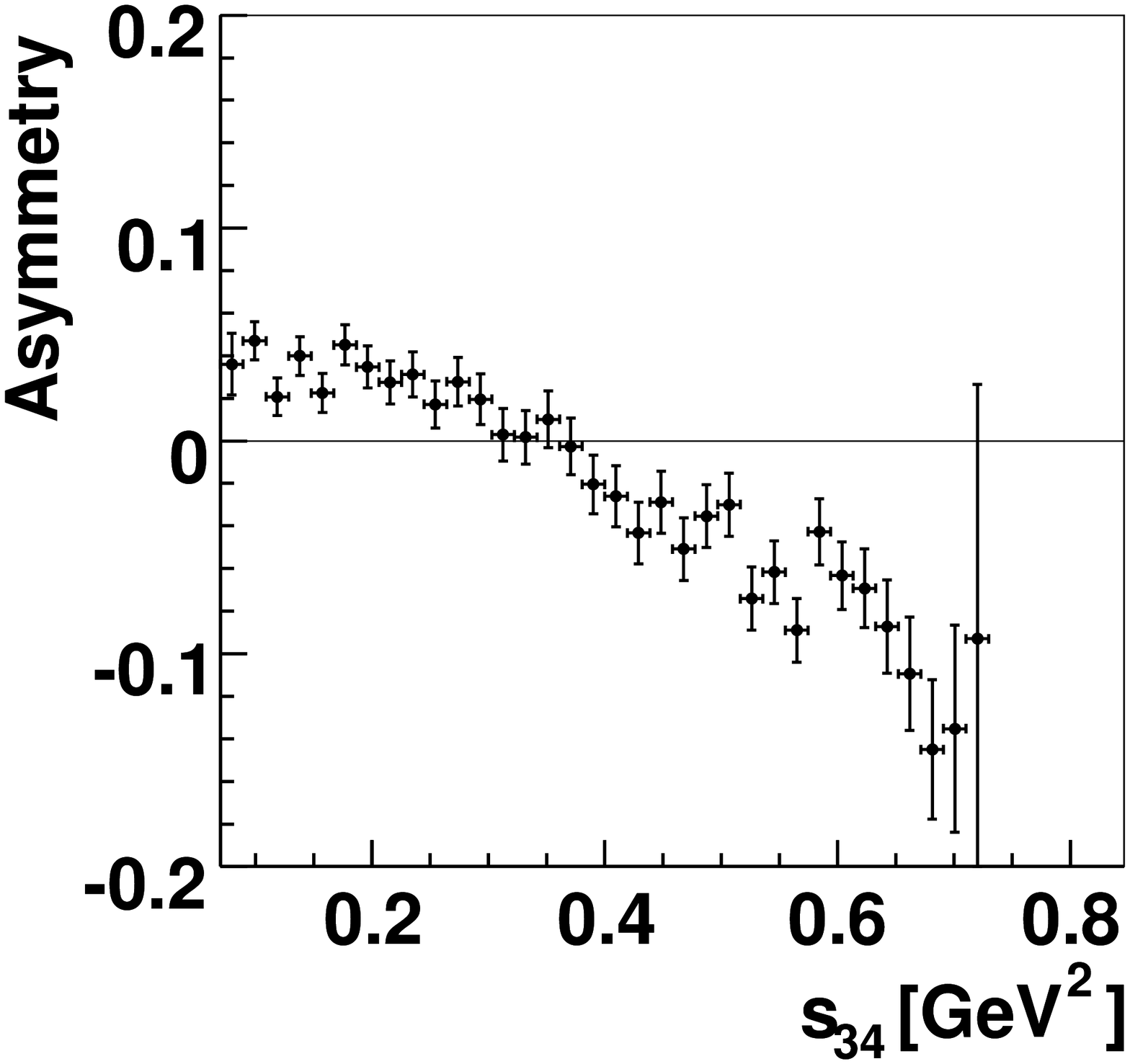,width=0.215\textwidth} \\
\epsfig{file=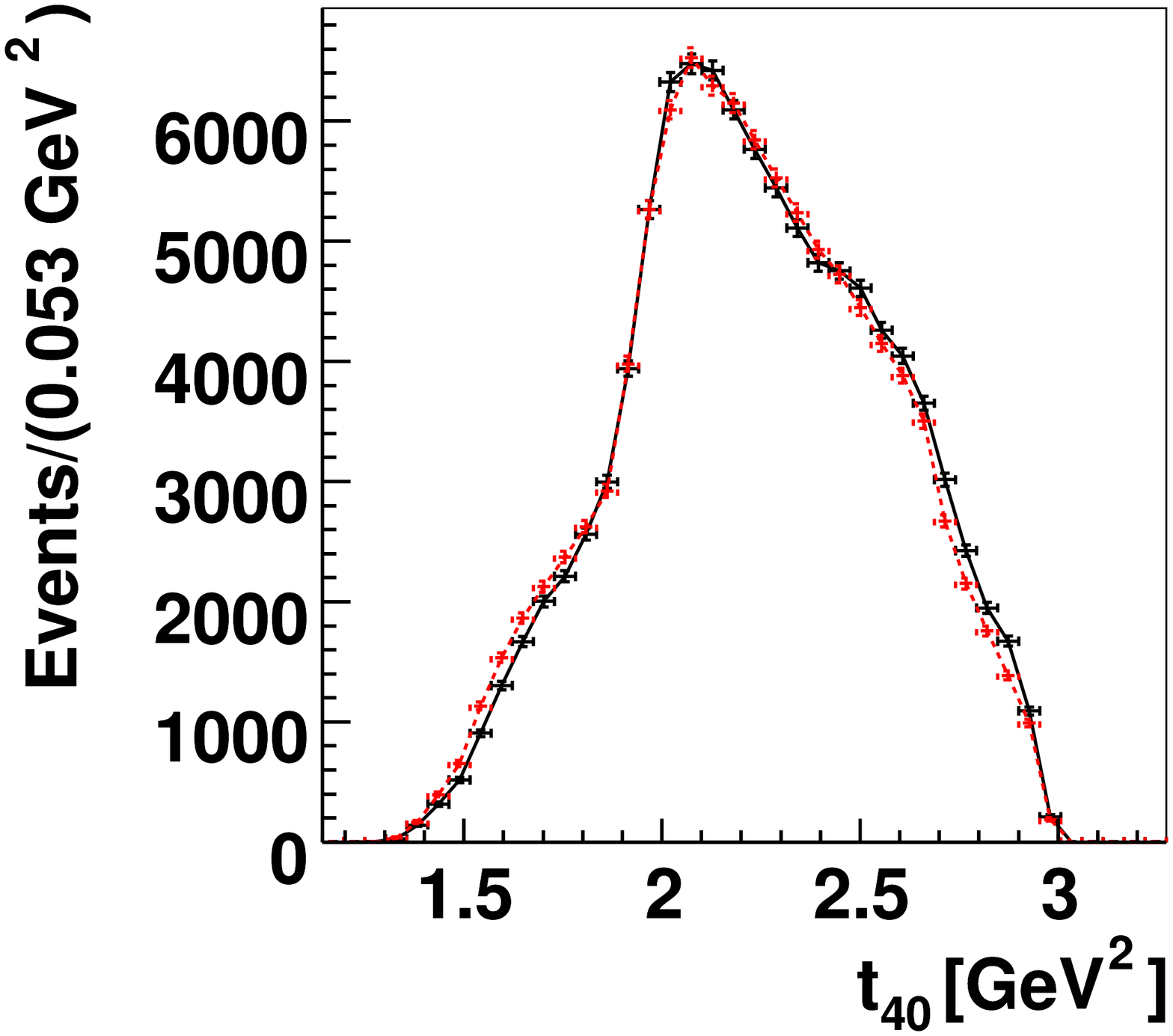,width=0.23\textwidth} & \epsfig{file=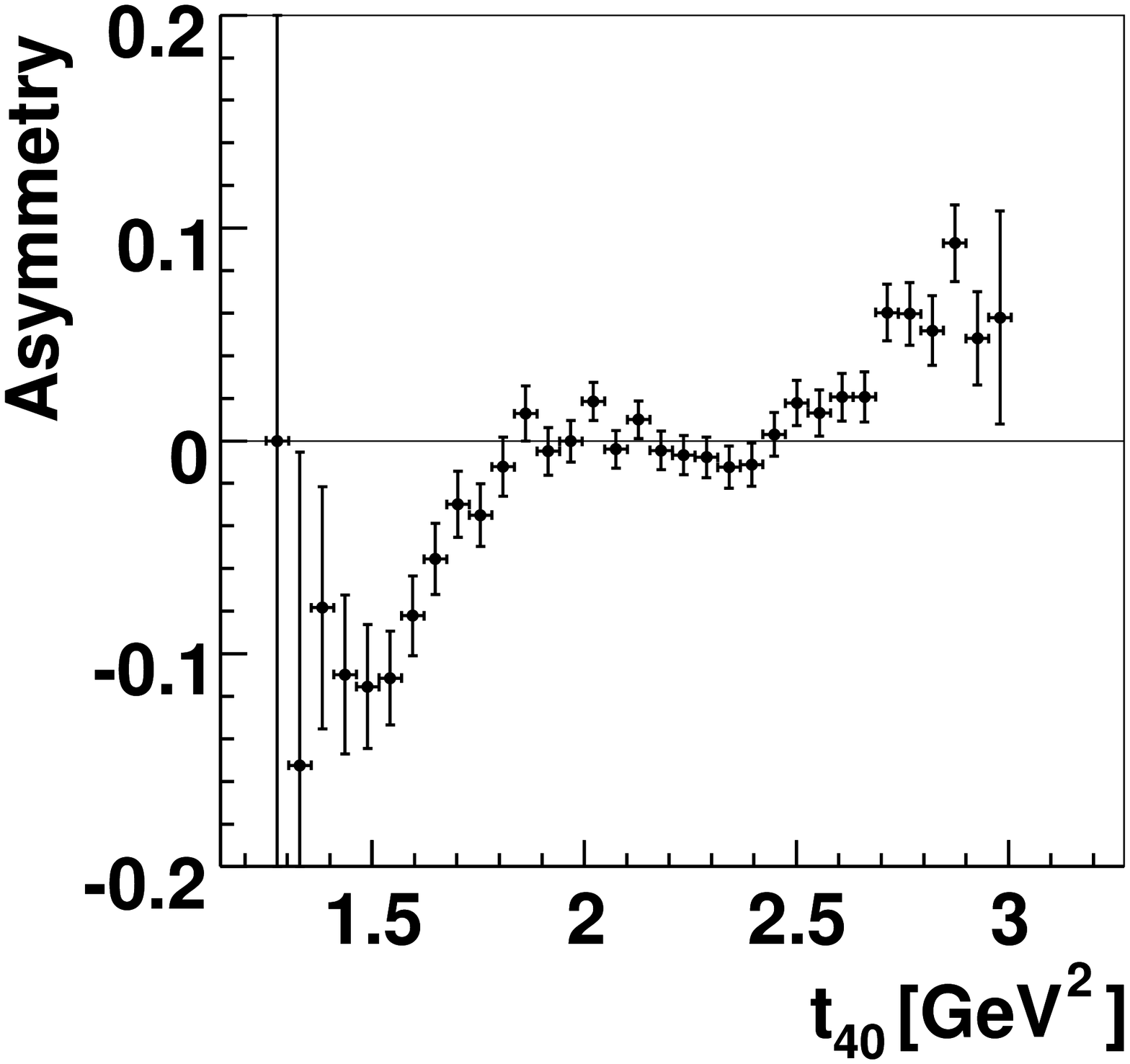,width=0.215\textwidth} 
\end{tabular}
\caption{Distributions (left) of 200k simulated events 
for the 5 kinematical variables shown for $\rm B^+$ (solid) and 
$\rm B^-$ (dashed) decays separately.  Also shown (right) are the asymmetries
between the $\rm B^+$ and $\rm B^-$ distributions, where the asymmetry
is defined as the number of $\rm B^-$ decays minus the number of $\rm B^+$
decays normalised by the sum.
\label{fig:tenk}
}
\end{center}
\end{figure}

To estimate the statistical precision achievable with this
method, we generated several simulation samples which
we then fitted to determine the parameters of interest,  most
notably $\gamma$. The samples were generated neglecting background 
and detector effects.  Figure~\ref{fig:tenk} shows the projections
of the chosen kinematical variables for 200k events, separately 
for  $\rm B^+$ and $\rm B^-$ decays, and the CP-asymmetry, 
defined as the number of $\rm B^+$ events minus 
the number of  $\rm B^-$ events, normalised by the sum.
In these projections the observable CP violation is small,
typically being at the few percent level only.
Full sensitivity to $\gamma$ is obtained through a likelihood fit
to all five variables.

\begin{figure}
\begin{center}
\epsfig{file=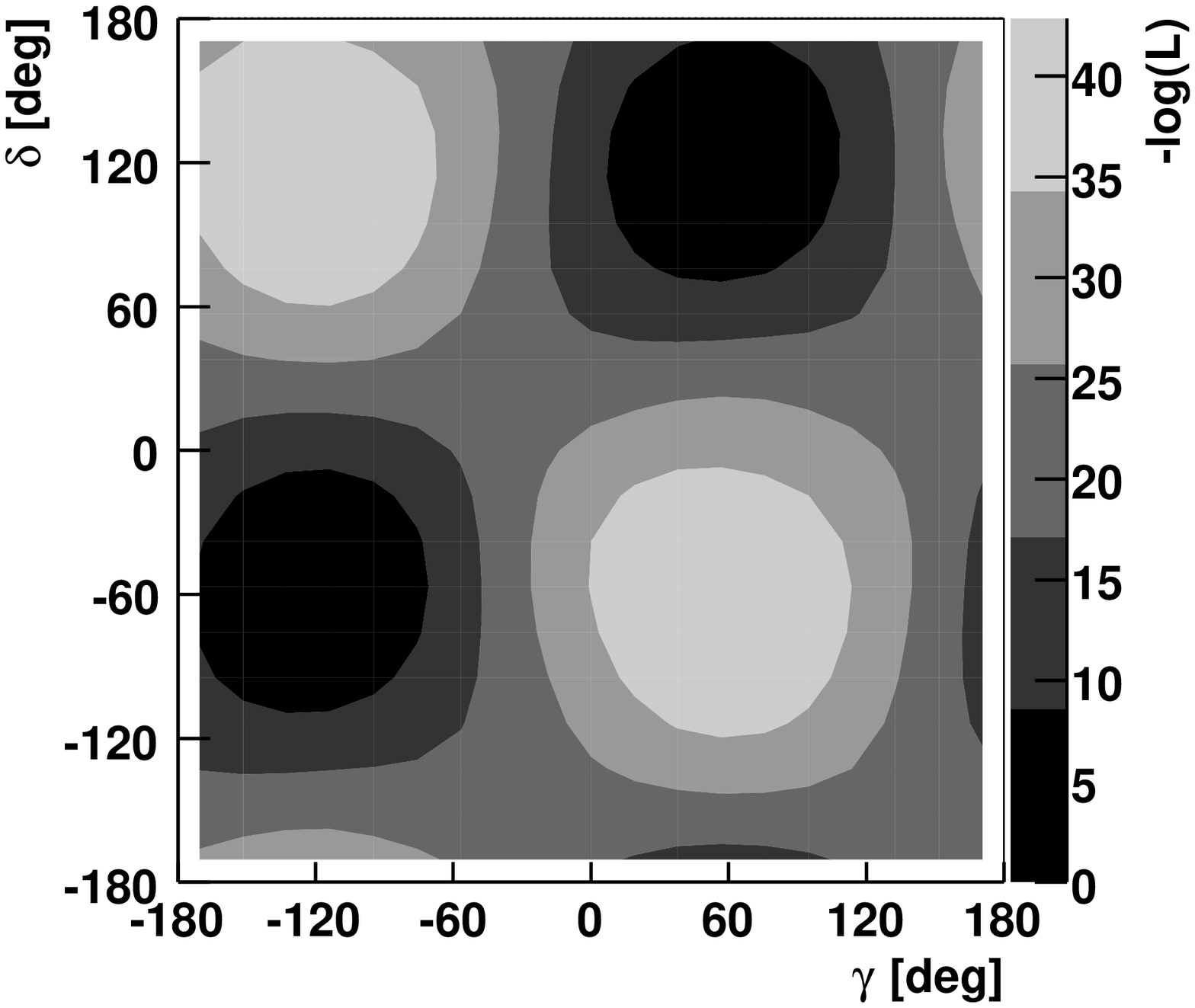,width=0.48\textwidth}
\epsfig{file=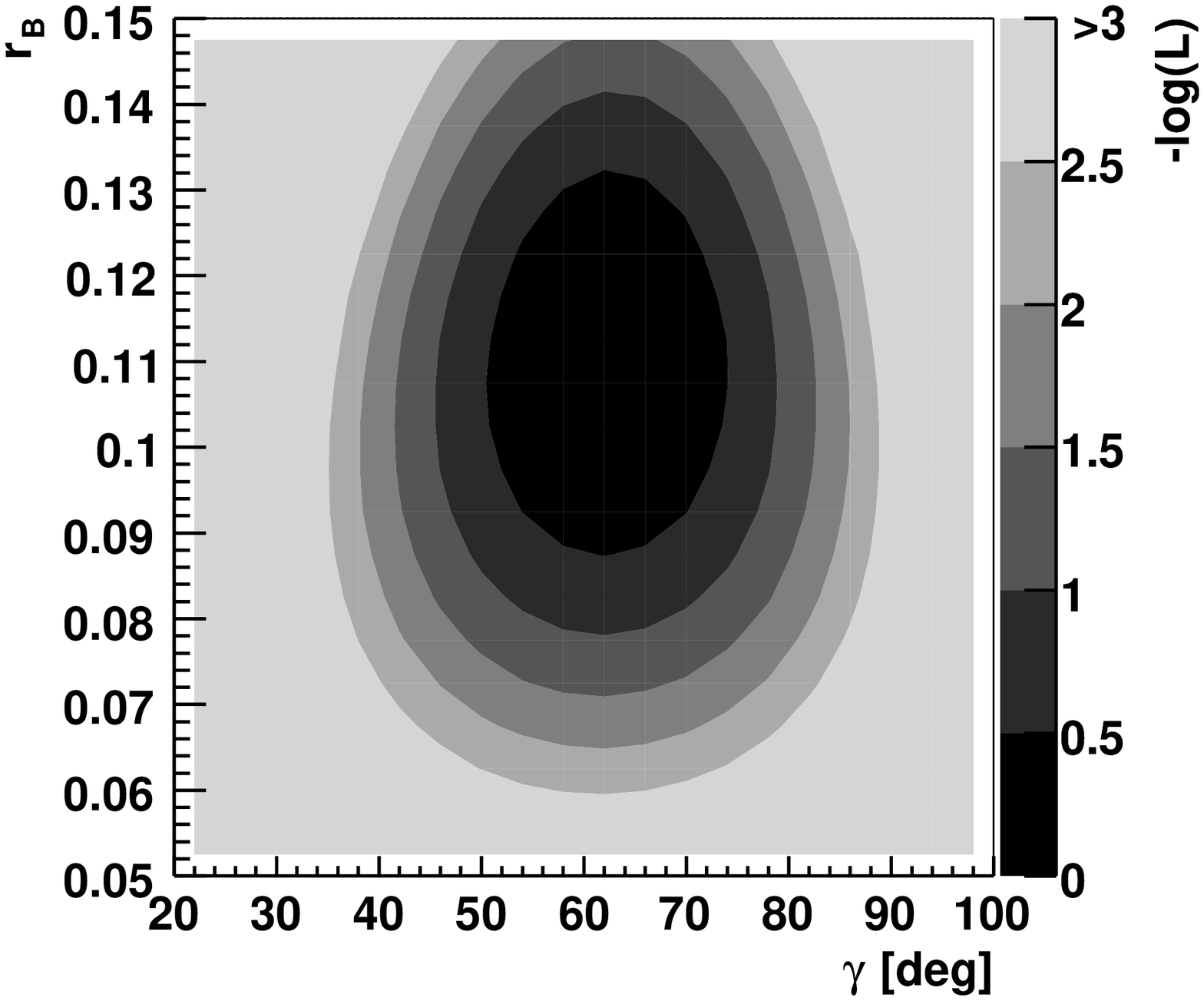,width=0.48\textwidth}
\caption{Negative log-likelihood shown for $\gamma$ against $\delta_B$ (top)
and $\gamma$ against $r_B$ (bottom) for a 
typical simulated experiment of 1000 events.   The input values
are at $\gamma=60^\circ$, $\delta_B=130^\circ$ and $r_B=0.10$.}
\label{fig:loglik}
\end{center}
\end{figure}

A log-likelihood function is defined as:
\begin{equation}
 \log(\mathcal{L})
 = 
 \sum\limits_{\mathrm{all\;B^-}} \log(P^-_i)
 + \sum\limits_{\mathrm{all\;B^+}} \log(P^+_j),
\end{equation}
where the probability density functions are defined as in 
expression~\ref{eq:pdf}, and the sums run over 
all B candidates in the sample.   The function was maximised
for each sample using the MINUIT package~\cite{MINUIT},  
with $r_B$, $\delta_B$ and $\gamma$ as the free parameters.
(It is assumed that all parameters associated with the D decay model
are known.)  Each sample contained 1000 events.
A scan of the negative log-likelihood, plotted for $\gamma$ against $\delta_B$ 
is shown for a typical sample in Figure~\ref{fig:loglik}.  The function
is well behaved with a minimum close to the input value and
a second solution at $\gamma - 180^\circ$ and $\delta_B - 180^\circ$.
Also shown is a scan for $\gamma$ against $r_B$.

For each sample the fitted parameters and the assigned errors
were recorded.  The reliability of the 
fit result was studied for each variable by 
constructing the `pull distribution', 
which is the difference between the fitted and input parameter,
divided by the assigned error.
The means and RMS widths of the pull
distributions are displayed in Table~\ref{tab:pull}
and are seen to be compatible with 0 and 1 respectively.
This indicates
that the log-likelihood fit is unbiased and the returned errors
are reliable.   The fit errors are also included in Table~\ref{tab:pull},
averaged over all fits.  $\gamma$ is extracted
with a precision of $14^\circ$.   There is very little correlation
between the three fit parameters, as is clear from the contours
in Figure~\ref{fig:loglik}.

\begin{table}
\begin{center}
\caption{Result from 218 simulated experiments, showing
the average assigned error, and the mean and width of the pull
distribution. \vspace*{0.2cm} \label{tab:pull}}
\begin{tabular}{ l  r@{$\pm$}r  r@{$\pm$}r  r@{$\pm$}r } \hline 
          &  \multicolumn{2}{c}{Error} & \multicolumn{2}{c}{Pull mean} & \multicolumn{2}{c}{Pull width} \\ \hline
 $\gamma$   & $14.4$ & $0.7^\circ$    & $0.13$ & $0.07$ & $1.10$ & $0.05$ \\
 $\delta_B$ & $14.3$ & $0.5^\circ$    & $0.10$ & $0.07$ & $1.09$ & $0.05$ \\
 $r_B$      & $0.023$ & $0.001$     & $0.11$ & $0.06$ & $0.95$   & $0.05$ \\
\end{tabular}
\end{center}
\end{table}

\begin{table}
\caption{Dependence of $\gamma$ fit results on the value of 
$r_B$, showing the average assigned error and the means and widths
of the pull distributions. \vspace*{0.2cm}  \label{tab:rB} }
\begin{center}
\begin{tabular}{ l r@{$\pm$}r  r@{$\pm$}r  r@{$\pm$}r }
\hline
\multicolumn{1}{c}{$r_B$}  &  \multicolumn{2}{c}{Error}  & \multicolumn{2}{c}{Pull mean} & \multicolumn{2}{c}{Pull width} \\ \hline
 $0.05$ &   $24.4$ & $0.6^\circ$  & $0.12$ & $0.21$   & $1.05$& $0.15$ \\
 $0.10$ &   $14.4$ & $0.7^\circ$  & $0.13$ & $0.07$   & $1.10$& $0.05$ \\
 $0.15$ &   $8.8 $ & $0.2^\circ$  & $0.00$ & $0.29$   & $0.96$& $0.21$ \\
 $0.20$ &   $7.2 $ & $0.1^\circ$  & $-0.06$& $0.22$   & $1.11$& $0.16$ \\
\end{tabular}
\end{center}
\end{table}

The size of the interference effects in \BgtoDK decays,  and
hence the sensitivity of the fit to $\gamma$,  depends on the value of $r_B$.
To investigate this dependence several 1000 event samples were generated  
with different values of $r_B$ between 0.05 and 0.20.  These samples
were then fitted as previously.  
The fit result on $\gamma$ and associated uncertainty for each $r_B$
value are shown in Table~\ref{tab:rB}.  It can be seen that the
$\gamma$ error varies approximately linearly with the inverse of $r_B$.


\begin{table}
\caption{Statistical uncertainty on $\gamma$ for various values of
$R$ and $\Delta \phi$.  These parameters are defined in 
expressions~\ref{eq:Rdef} and~\ref{eq:Dpdef} with the same values
being used for $\rm  K_1(1270)K$, $\rm K_1(1270)K$ and 
$\rm K^\star (892)^0 K \pi$.
\vspace*{0.2cm}
\label{tab:scenario1}}
\begin{center}
\begin{tabular}{lcccc}
\hline
 $R$  & \multicolumn{4}{c}{$\Delta \phi$} \\ \cline{2-5}
      &  $0^\circ$ & $90^\circ$ & $180^\circ$ & $270^\circ$ \\ \hline
$    0$     &  $14^\circ$ &  / & /  &  / \\
$ 0.25$     &  $18^\circ$ & $14^\circ$ & $19^\circ$ & $13^\circ$ \\
$ 0.50$     &  $27^\circ$ & $13^\circ$ & $18^\circ$ & $13^\circ$ \\
$ 1.00$     &  $23^\circ$ & $13^\circ$ &    /     & $13^\circ$ \\
$ 2.00$     &  $21^\circ$ & $14^\circ$ & $19^\circ$ & $14^\circ$ \\ 
\end{tabular}
\end{center}
\end{table}

\begin{table}
\caption{Statistical uncertainty on $\gamma$ for various $R$ values 
and different $\Delta \phi$ scenarios. 
The same 
$R$ values are
being used for $\rm  K_1(1270)K$, $\rm K_1(1400)K$ and 
$\rm K^\star (892)^0 K \pi$.
The scenarios for $\Delta \phi$ are given in the text. \vspace*{0.2cm}
\label{tab:scenario2}}
\begin{center}
\begin{tabular}{lcccc}
\hline
 $R$  & \multicolumn{4}{c}{$\Delta \phi$ scenario} \\ \cline{2-5}
      &  1 & 2 & 3 & 4 \\ \hline
$ 0.25$     &  $14^\circ$ & $19^\circ$ & $14^\circ$ & $13^\circ$ \\
$ 0.50$     &  $14^\circ$ & $17^\circ$ & $14^\circ$ & $15^\circ$ \\
$ 1.00$     &  $12^\circ$ & $18^\circ$ & $26^\circ$ & $20^\circ$ \\
$ 2.00$     &  $13^\circ$ & $14^\circ$ & $16^\circ$ & $19^\circ$ \\ 
\end{tabular}
\end{center}
\end{table}

As explained in Section~\ref{sec:model} the fitted model reported
in~\cite{FOCUS} does not distinguish between the relative 
contribution of certain $\rm D$ decay amplitudes and their
CP-conjugate final states. The importance of this unknown information
on the fit sensitivity was assessed by generating and fitting
1000 event simulated datasets with different
values of the $R$ and $\Delta \phi$ parameters defined
in expressions~\ref{eq:Rdef} and~\ref{eq:Dpdef}.  In varying these
parameters the overall contribution of each mode and its CP-conjugate
state, eg.  $|A({\rm D^0 \to K_1(1270)^{+}K^{-}})\, + \,
A({\rm D^0 \to K_1(1270)^{-}K^{+}})|^2$, was kept constant.
The results
for the uncertainty on $\gamma$ are 
shown in Table~\ref{tab:scenario1} in the case where 
a common value of $R$ and $\Delta \phi$ is taken for the three
final states under consideration.
In a further study the phase shift $\Delta \phi$ was allowed
to take different values between the three modes.  Four scenarios
were considered with the following arbitrary (randomly chosen) sets
of values for $\Delta \phi_{K_1(1270)K}$,  $\Delta \phi_{K_1(1400)K}$
and $\Delta \phi_{K^\star (892)^0 K \pi}$ respectively:
\begin{enumerate}
\item{39$^\circ$, 211$^\circ$ and 115$^\circ$ (default);}
\item{53$^\circ$, 108$^\circ$ and 15$^\circ$;}
\item{55$^\circ$, 344$^\circ$ and 173$^\circ$;}
\item{209$^\circ$, 339$^\circ$ and 87$^\circ$.}
\end{enumerate}
\noindent The statistical uncertainties found on $\gamma$ for these scenarios
are given in Table~\ref{tab:scenario2}.  For both 
Table~\ref{tab:scenario1} and Table~\ref{tab:scenario2} only 
a single experiment was performed at each point in parameter space, hence
the stated error carries an uncertainty of a few degrees. However any 
minor variation in result arising from the exact value of 
the fitted $r_B$ parameter, experiment-to-experiment,  has been
corrected for by using the dependence observed in the study reported in
Table~\ref{tab:rB}.   

It can be seen that the precision of the fit is fairly uniform
over parameter space,  with a typical value of $15^\circ$. In certain
cases however the precision is worse,  particularly when $R=1$ and/or
$\Delta \phi = 0$.
More detailed studies of \Dgto2k2p decays are therefore needed 
to reliably estimate the instrinsic sensitivity of \BbtoD2k2pK 
for a $\gamma$ measurement.  However, variations in other aspects
of the \Dgto2k2p 
decay structure were found to have limited consequences for the
fit precision.

Finally it was investigated what biases would be introduced in
the $\gamma$ extraction through incorrect knowledge of the decay
model.  Experiments were performed in which the datasets were
generated with the full model in the default scenario,  but 
fitted with a model which omitted all the decay amplitudes
with a contribution less than 3\% to the overall rate.   Shifts
of up to $8^\circ$ were observed in the measured value of $\gamma$.
This value can be considered
as an upper bound to any final systematic uncertainty,  as it will
be possible to accumulate very large samples of \Dgto2k2p events
at the LHC,  which will allow the decay model to be refined and improved
with respect to the one assumed here.  Additional information will also
become available from CP-tagged $\rm D$ decays at
facilities operating at the $\psi (3770)$ resonance~\cite{BONDAR}.

\section{Conclusions}
\label{sec:conc}

We have shown that the decay \BbtoD2k2pK  can be used to provide
an interesting measurement of the unitarity triangle angle $\gamma$.  
With 1000 events and assuming a value of $r_B=0.10$ it is possible
to measure $\gamma$ with a precision of around $15^\circ$.  
The exact sensitivity achievable depends on the relative contributions
of certain unmeasured modes in the $\rm D$ decay model.
The final state, involving only charged particles, and kaons in particular,
is well suited to LHCb.  A full reconstruction study is necessary to
estimate reliably  the expected event yields and the level of background.

Finally we remark that the same technique of a four-body amplitude
analysis in \BgtoDK decays can be applied to other modes,  most
notably the `ADS' channel $\rm D \to K^\pm \pi^\mp \pi^+ \pi^-$.

\vspace*{0.3cm}
\section*{Acknowlegements}

We are grateful to David Asner, Robert Fleischer and 
Alberto Reis for valuable discussions. 
We also acknowledge the support of the  
Particle Physics and Astronomy Research Council,  UK.

\end{document}